\documentclass[prl,twocolumn,amsmath,nobibnotes,nofootinbib,superscriptaddress,preprintnumbers]{revtex4}

\usepackage{bm} \usepackage{graphicx} 
\usepackage{epstopdf}
\def\ba{\begin{eqnarray}} \def\ea{\end{eqnarray}}
\def\be{\begin{equation}} \def\ee{\end{equation}} \def\({\left(}
\def\){\right)} \def\[{\left[} \def\]{\right]} \def\<{\left<}
\def\>{\right>}

\def\ba{\begin{eqnarray}}
\def\ea{\end{eqnarray}}
\def\be{\begin{equation}}
\def\ee{\end{equation}}
\def\({\left(}
\def\){\right)}
\def\[{\left[}
\def\]{\right]}
\def\<{\left<}
\def\>{\right>}

\begin{document}

\title{An obstacle to populating the string theory landscape}
\date{\today}

\author{Matthew C Johnson}
\affiliation{California Institute of Technology, Pasadena, CA 91125, USA} 
\author{Magdalena Larfors}
\affiliation{Institutionen f\"or Fysik och Astronomi, Box 803, SE-751 08 Uppsala, Sweden. }

\begin{abstract}
We construct domain walls and instantons in a class of models with coupled scalar fields, determining, in agreement with previous studies, that many such solutions contain naked timelike singularities. Vacuum bubble solutions of this type do not contain a region of true vacuum, obstructing the ability of eternal inflation to populate other vacua. We determine a criterion that potentials must satisfy to avoid the existence of such singularities, and show that many domain wall solutions in Type IIB string theory are singular. 
\end{abstract}

\preprint{UUITP-21/08, CALT-68.2700}

\maketitle

{\em Introduction}: The string theory landscape (a set of vacua in the low--energy approximation of string theory) has changed the way we think about making predictions from fundamental theory. Together with eternal inflation (the idea that a background de Sitter space can form many different pocket universes but still perdure) as a mechanism to populate the various states in the landscape, it becomes possible to think of many physical quantities as environmental variables that vary over vast spatiotemporal regions of a very large "multiverse." In this picture, one can only make statistical predictions based on dynamical criteria or selection effects, with the most striking (and successful) example being a prediction for the value of the cosmological constant~\cite{Weinberg:1987dv}. 

In order to make sense of a probability measure over different vacua, it is necessary to understand how exactly eternal inflation populates the landscape. This is typically assumed to proceed via the formation of vacuum bubbles containing different phases, a process described by the Coleman--de Luccia (CDL) instanton~\cite{Coleman:1980aw}. The potential that forms the vacuum bubble is often taken to depend on a single scalar field only. However, it was shown by Cvetic and Soleng~\cite{Cvetic:1994ya} that domain wall spacetimes in certain two--field models contain naked singularities (see also Ref.~\cite{Saffin:1998kg} for a similar observation in a slightly different context). Such singularities prevent the formation of regions of a new phase inside of vacuum bubbles and might forbid certain vacuum transitions~\cite{Vilenkin:1998pp}. In this paper, we show that the best--understood corner of the landscape, described by Type IIB string theory, gives rise to many such singular 
 domain wall solutions. The lack of standard instantons and domain walls pose a serious concern about the ability to connect vacua and make statistical predictions based on eternal inflation.

{\em Singular domain walls and instantons}: Consider the construction of domain walls and instantons in an effective four--dimensional theory of two real scalar fields $\{\phi, \chi\}$ coupled to gravity, described by the Lagrangian (we use units where $M_p=1$)
\begin{eqnarray}\label{eq:action}
\mathcal{L} = -\frac{1}{16 \pi} R + \partial_{\mu} \chi \partial^{\mu} \chi + \partial_{\mu} \phi \partial^{\mu} \phi - V_0 (\phi, \chi) - V_1 (\chi),
\end{eqnarray}
where  the potential $V_0(\phi,\chi)$ is given by
\begin{equation}\label{eq:Vphi}
V_0 (\phi,\chi) = \mu^4 f(\chi) \left[ \left( \left( \frac{\phi}{M} \right)^2 - 1 \right)^2 - \alpha \frac{\phi}{M} + C \right].
\end{equation}
The constants $\alpha$ and $C$ are chosen such that $V_0$ has two positive minima located at $\phi_{\pm} \sim \pm M$. The dimensionless function $f(\chi)$ modulates the height of the potential barrier between the two minima (thus $\chi$ acts as a dilatonic field). We focus on functions $f(\chi)$ which diverge to $+\infty$ as $\chi$ goes negative and monotonically approach zero from above as $\chi \rightarrow \infty$, such as $f(\chi) \propto \exp(- n \chi)$ ($n \ge 0$). The potential $V_1 (\chi)$ contains self--interaction terms for the field $\chi$, which we assume can stabilize the field at a positive VEV, yielding two full vacua at $\phi_{\pm}$.

For $\alpha \neq 0$, there can exist closed domain wall solutions (bubbles) interploating between $\phi = \phi_{\pm}$. Their nucleation is described by the CDL instanton~\cite{Coleman:1980aw}, which for two positive energy minima, is a compact Euclidean manifold covered by the metric
\be
ds^2 = d\xi^2 + r(\xi)^2 d \Omega_3^2.
\ee
Analytically continuing one of the angular variables of the three--sphere $d\Omega_3$ yields the post--nucleation metric near the domain wall, whose sections of constant $\xi$ are hyperboloids with spacelike norm. Using either the domain wall or CDL metric ansatz, the equations of motion are
\begin{equation}\label{eq:eoms}
\phi'' + 3\frac{r'}{r} \phi' = \frac{\partial V_0 }{\partial \phi}, \ \ \chi'' + 3 \frac{r'}{r} \chi' = \frac{\partial V_0}{\partial \chi} + \frac{\partial V_1}{\partial \chi},
\end{equation}
\begin{equation}
r'^2 = 1 + \frac{8 \pi r^2}{3 } \left[ \frac{\phi'^2}{2} + \frac{\chi'^2}{2} - V_0 - V_1 \right].
\end{equation}
To find non--singular interpolating solutions between the basins of attraction of $V_0$, we must solve a double boundary value problem at the poles, $\xi=0$ and $\xi=\xi_f$, of the compact spatial slices. Specifically, $\{ \phi (0) \sim \phi_-, \phi' (0) = \chi' (0) = 0,  r(0) = 0, \phi(\xi_f) \sim \phi_+, \phi' (\xi_f) = \chi' (\xi_f) = 0, r(\xi_f) = 0 \}$. We will often consider the mechanical analog of a particle moving as a function of the "time" $\xi$ in the upside down Euclidean potential.

When $\mu^4 \ll 1$, $M < M_p$, and $\alpha \ll 1$, we can take what is known as the thin--wall limit~\cite{Coleman:1980aw}. In this case, the field spends most of its time loitering very near both the true and false Euclidean maxima, and crosses between the two essentially instantaneously. The smallness of the vacuum energies also implies that the scale factor increases approximately linearly over a significant range in $\xi$.

In a theory with just $\phi$, the construction of non--singular instantons and their associated domain wall spacetimes is well known, and can be accomplished analytically in the thin--wall approximation~\cite{Coleman:1980aw}. However, the addition of $\chi$ can drastically alter the structure of domain wall solutions, as first noted in \cite{Cvetic:1994ya}, see also~\cite{Cvetic:1995rp,Saffin:1998kg}. 

\begin{figure}
\includegraphics[scale=.8]{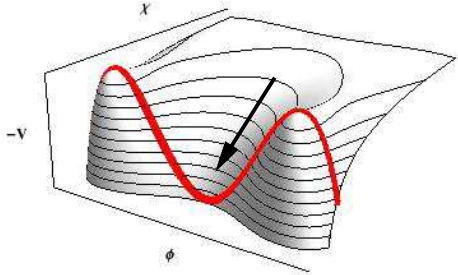}
\caption{
 \label{fig:potential}
A Euclidean potential that leads to singular domain wall and instanton solutions. The arrow indicates the strong gradient towards $-\chi$ that exists away from vacuum.}
\end{figure}

To illustrate this, consider the case where $V_0$ is negligible for $\phi \simeq \phi_{\pm}$, but $V_0 \gg V_1$ when $\phi$ is displaced away from either minimum. An example of such a Euclidean potential is shown in Fig.~\ref{fig:potential}. The presence of the function $f(\chi)$ induces a significant gradient in the {\em negative}--$\chi$ direction whenever $\phi$ is away from its minima. The velocity gained by $\chi$ as $\phi$ makes its traverse between the vacua can be calculated by integrating the equation of motion for $\chi$, assuming that $\chi$ is fixed at $\chi_0$ (a good approximation for an instantaneous jump in $\phi$). The friction term is negligible during this stage of the evolution (since $r \propto \xi$ and $\xi \gg 1$ here), and we obtain (in agreement with~\cite{Cvetic:1994ya})
\begin{equation}\label{eq:deltachi'}
\Delta \chi' = \left( \frac{\partial (\ln f)}{\partial \chi} \right)_{\chi_0}  \int_{\xi_w - \epsilon}^{\xi_w + \epsilon} d\xi V_0 = \frac{\sigma}{2} \left( \frac{\partial (\ln f)}{\partial \chi} \right)_{\chi_0}.
\end{equation} 
with $\epsilon \rightarrow 0$. The last equality defines the tension $\sigma$ of the domain wall.

This gained velocity is very dangerous. If $\chi'$ is non-zero as $r$ approaches its second zero, the neglected (anti) friction term in the equations of motion will become significant, sending $\chi' \rightarrow - \infty$ in finite $\xi$, and the result is a curvature singularity~\cite{Cvetic:1994ya}. This catastrophe can only be avoided if the potential $V_1$ can halt the motion of the $\chi$ field. Since energy is approximately conserved when $\phi$ makes its traverse between the vacua, this will only be possible when $V_1$ has a local maximum at $\chi = \chi_{max}$ with 
\begin{equation}\label{eq:instcondition}
V_1 (\chi_{max}) > \Delta \chi'^2 / 2 
\end{equation}
at both $\phi = \phi_{\pm}$. In this case, bound trajectories for $\chi$ in the Euclidean potential can exist, and non--singular solutions can possibly be found (this condition is necessary, but not sufficient). 

\begin{figure}
\includegraphics[scale=0.33]{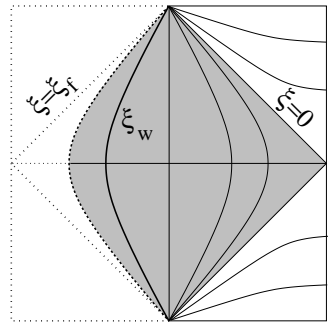}
\caption{
  \label{fig:conformald}
The time--symmetric thin-wall singular vacuum bubble spacetime. The shaded region is described by the solution to Eq.~\ref{eq:eoms}. 
The bubble wall is the thick solid line at $\xi_w$, and the singularity is the dashed line (beyond which the non--singular solution would have relaxed to the true vacuum). 
}
\end{figure}

The appearance of the singularity has profound implications for the structure of the domain wall spacetime, as shown in Fig.~\ref{fig:conformald}. If we choose to set $\chi' = 0$ on the false vacuum side of the potential, then we can extend the coordinates across $\xi = 0$ to a region where the fields settle to the false vacuum. In the reglular CDL solution it is also possible to extend the coordinates across $\xi = \xi_f$ (on the other side of the domain wall) to a region where the fields settle to the true vacuum. However, in the singular domain wall spacetimes, {\em this region does not exist} -- there is no region of true vacuum!

Cvetic and Soleng \cite{Cvetic:1994ya} showed that in addition to the CDL bubbles discussed here (a.k.a. ultra-extreme domain walls), spacetimes containing extreme and non-extreme domain walls in theories described by Eq.~\ref{eq:action} will also exhibit singular behavior.  We expect this to be a generic phenomenon whenever the fields are a function of a single spacelike variable (i.e. $\xi$ in the case of CDL bubbles and physical radius in the case of static $O(3)$--symmetric vacuum bubbles~\cite{Garriga:2004nm}). 

Singular instantons (of finite action~\cite{Hawking:1998bn,Saffin:1998kg}) have also been considered previously~\cite{Hawking:1998bn}, as describing the birth of an open universe in the absence of an antecedent false vacuum. However, when the singularity lies on the true vacuum side of the instanton (as above), it has been argued that such finite action instantons cannot be interpreted as an instability of the false vacuum~\cite{Vilenkin:1998pp}. Thus, CDL transitions between the vacua at $\phi \sim \pm M$ would seem to be forbidden. (Note that there is nothing forbidding domain walls and instantons developing between vacua in the $\chi$ direction while $\phi$ is fixed in vacuum.)

Let us now relax some of our assumptions, and ask when non--singular solutions could exist. Keeping with the thin--wall approximation, unless $V_0 \sim V_1$ everywhere between the minima, Eq.~\ref{eq:deltachi'} holds, and $\chi$ obtains a negative velocity. Thus, even when the condition Eq.~\ref{eq:instcondition} is satisfied, we may have to consider potentials where the VEVs of $\chi$ in the two minima are very finely tuned according to the velocity gained. Alternatively, if the fields and metric are allowed to depend both on a timelike and a spacelike variable, say by requiring $O(3)$ invariance alone, singularities could  possibly be avoided. To analyze the behavior of these solutions lies beyond the scope of this letter, but certainly deserves further investigation.

Moving away from thin--wall, and considering potentials where gravitational effects become important, the friction term in the field equations can no longer be neglected, but clearly the trend will still be to push $\chi$ to zero. When gravitational effects dominate, it becomes impossible to find a CDL solution even in the case where there is a single field. In this regime, the Hawking Moss (HM) instanton~\cite{Hawking:1981fz} (where all fields sit at an extremal point of the potential over the entire range in $\xi$) can mediate a transition. However, because there is no extremal point between $\phi_{\pm}$ (at the maximum in $\phi$, $\partial V / \partial \chi \neq 0$), this class of potentials will not admit a HM instanton (see also ~\cite{Johnson:2008kc}). The lack of a solution both in the limit where gravitational effects are negligible and where they are dominant is strong evidence for the lack of solutions in the intermediate regime.

However, if $V_0$ is everywhere comparable to $V_1$, then our conclusions can change appreciably. In this case, an extremal point can exist between the minima, allowing a HM instanton, and providing a location where the force pushing $\chi$ to zero in the CDL solution disappears. In general, precise statements are difficult to make, but we would expect that non--singular solutions exist in a wide variety of such potentials (for example, see~\cite{Saffin:1998kg}).

{\em Singular domain walls in the string theory landscape:} Like all string/M theories, type IIB can be compactified on an internal manifold, yielding a lower dimensional effective field theory at low energies. Choosing the internal manifold to be a Calabi--Yau three-fold, yields a four-dimensional supergravity (provided that the internal volume is large in string units and that the string coupling is small). Extra ingredients, such as higher-dimensional fluxes and branes, are needed to stabilize the geometric moduli of the Calabi--Yau (see e.g. \cite{Douglas:2006es} for reviews).

In type IIB flux compactifications, the complex structure moduli $z_i$ and the axio-dilaton $\tau$ receive a tree level superpotential, $W_{0}(z_i,\tau; F,H)$, determined by quantized three-form fluxes $F$ and $H$. The K\"ahler moduli $\rho_a$, on the other hand, can only be fixed by non-classical effects. For simplicity, we restrict to the case with one K\"ahler modulus $\rho$, whose imaginary part determines the overall volume of the Calabi--Yau: $\mbox{Vol} \propto \rho_I^{3/2}$. 

The tree-level superpotential $W_0$, combined with the K{\"a}hler potential $K = -3 \log (\rho_I) - \log (\tau_I) + K_{cs}(z_i, \bar{z_i})$, yields an $\mathcal{N}=1$ scalar potential at tree level:
\begin{eqnarray}
V_T &=& \frac{e^{K_{cs}}}{\rho_I^3 \tau_I} \left( K^{z_i \bar{z_i}} D_{z_i} W_{0} D_{\bar{z_i}} \overline{W}_{0} \nonumber  + K^{\tau \bar{\tau}} D_{\tau} W_{0} D_{\bar{\tau}} \overline{W}_{0} \right).
\end{eqnarray}
$V_T$ acts to stabilize $z_i$ and $\tau$ and is positive definite, but induces a run-away behavior for $\rho$. For fixed $z_i$, $V_T$ has a global minimum determining the string coupling $g_s = <\tau_I>$ ~\cite{Danielsson:2006xw}. It also has a minimum in complex structure moduli space for generic flux configurations. 

The superpotential receives non-perturbative corrections of the form $W_{np}(\rho) = A e^{i a \rho}$ \cite{Witten:1996bn, Kachru:2003aw} that can fix the K\"ahler modulus $\rho$. This effect has been shown to give viable vacua in the supergravity regime, as described by the KKLT model~\cite{Kachru:2003aw} (see Ref.~\cite{Balasubramanian:2005zx} for another approach), which we focus on here.

The set of vacua labeled by different flux configurations is known as the string theory landscape~\cite{Susskind:2003kw}. Vacua with different flux configurations can be connected by a continuous potential $V_T$ using the monodromies of the underlying Calabi--Yau~\cite{Danielsson:2006xw,Chialva:2007sv}, yielding a double--well potential~\cite{Johnson:2008kc} not unlike $V_0$ of the previous section. Freezing $\rho$, there are domain wall solutions in this potential~\cite{Johnson:2008kc}, which in the thin wall approximation, have tension 
\begin{equation}
\sigma = h g_s^{1/2} \rho_I^{-3/2}  ,
\end{equation}
where $h$ is a dimensionless number depending on the instanton path in moduli space~\cite{Johnson:2008kc}. The same tension is also obtained from wrapped D and/or NS five-branes \cite{Feng:2000if,Frey:2003dm,deAlwis:2006cb,Freivogel:2008wm}. In that picture, $h$ is determined by the tension of the five-brane and the volume of the cycle the brane is wrapping.

Assuming that $W_0$ is real, and neglecting $\rho_R$ (which will receive a mass), the K\"ahler modulus potential $V_{\rho}$ is
\begin{equation}
V_{\rho} = \frac{a A e^{- a \rho_I}}{2 \rho_I^2} \left[ W_0 + A e^{-a \rho_I} \left( 1 + \frac{1}{3}a \rho_I \right) \right] + \frac{D}{\rho_I^3},
\end{equation}
where $a$, $A$, and $D \geq 0$ are undetermined (but in principle calculable) constants~\cite{Kachru:2003aw}. For vacua where supersymmetry is broken by $\rho$ alone, and $W_0$ is negative and small in magnitude, it is possible to find a positive energy minimum at $\rho_I \gg 1$. The height of the barrier separating the minimum from the run--away part of the potential at $\rho_I \rightarrow \infty$ can be approximated by~\cite{Kallosh:2004yh} 
\begin{equation}\label{eq:bheight}
V_{\rho} (\rho_{max}) \simeq \frac{a^2 A^2 e^{-2 a \rho_I}}{6 \rho_I}.
\end{equation}

Assuming that two neighboring vacua  of $V_{T}$ with the appropriate values of $W_0$ can be found (this is likely to be a very rare set of minima, see~\cite{Dine:2008jx, Johnson:2008kc}), this completes the analogy with the model of the previous section, with $V_{\rho}$ playing the role of $V_1$. At large volume, where the effective theory is valid, $V_{T} \gg V_{\rho}$ everywhere except in the very near vicinity of the vacua for $\{z,\tau\}$.

When a transition between flux vacua occurs, the K\"ahler modulus will
get a kick just as in our toy model. Since the K{\"a}hler metric is
block diagonal in the $z_i$, $\tau$ and $\rho$ sectors,
we can go to the basis where $\rho$ has a canonical kinetic term. 
Performing the transformation $\rho_I = e^{\sqrt{2/3} \chi
}$, Eq.~\ref{eq:deltachi'} can now be used to estimate the
kinetic energy gained across the wall, with $V_T$ playing the role of
$V_0$ and $f(\chi) = e^{-\sqrt{6} \chi }$. The change in
kinetic energy incurred by $\chi$ across the wall can then be compared with the barrier height Eq.~\ref{eq:bheight}, 
\begin{equation}
\frac{\chi'^2 / 2}{V_{\rho} (\rho_{max})} \simeq \frac{9 g_s h^2}{a^2 A^2} \frac{e^{2a\rho_I}}{\rho_I^2} \gg 1.
\end{equation}
Thus, at large volume, the kinetic energy is naturally many orders of magnitude larger than the barrier height! As an example, for the stabilization parameters in~\cite{Kachru:2003aw} and a typical tension found in the Mirror Quintic model of~\cite{Johnson:2008kc}, $\{A=1, a=.1, g_s^{1/2}h=.1, \rho_I \sim 100 \}$, this yields a fraction on the order of $10^6$. This conclusion is directly related to the different scales of $V_T$ and $V_{\rho}$ at large volume, and would seem to hold in other models for moduli stabilization in type IIB, such as~\cite{Balasubramanian:2005zx}. 

Just as in the toy model, this conclusion may be avoided if $V_T \sim V_{\rho}$. This can happen when the Calabi--Yau volume is not large, but this means that we leave the supergravity regime. Another possibility is if the string coupling $g_s$ and/or $h$ (as determined by $V_T$) could be as small as $\mathcal{O}(\rho_I^2 e^{-2a\rho_I})$. If the minima of $V_T$ lie very near a conifold point in the complex structure moduli space, then $h$ could in principle be small as discussed in \cite{Frey:2003dm,Johnson:2008kc,Freivogel:2008wm}. Similarly, tuned fluxes could fix $g_s = <\tau_I>$ at a suitable value.

If gravitational effects are important, we might expect to find thick-wall CDL and HM instantons mediating the tunneling between vacua. The thick-wall CDL instantons found in~\cite{Johnson:2008kc} are more difficult to analyze than the thin-wall instantons discussed here, but we believe the disparity in scales among the moduli sectors will produce similar effects. Furthermore, just as in the toy model, there are no HM points in the type IIB potential, as long as the scales of $V_T$ and $V_{\rho}$ are disparate. 

{\em Conclusions and future directions:} We have found that, because of the disparity in scales fixing
different moduli, many of the domain wall solutions and instantons in
the Type IIB landscape are singular. This creates an obstacle for eternal inflation to populate regions of the landscape. The breadth of these conclusions in the space of Type IIB theories certainly deserves further investigation. In particular, it would be interesting if only distinguished states
participate in eternal inflation. It is also important to search for
corners of the landscape where the form of the potential might be
different. For example, in Type IIA theories it is possible to
stabilize all moduli at tree level with flux~\cite{DeWolfe:2005uu}, avoiding a hierarchy in
scales in the potential fixing the various moduli, and thus
potentially avoiding singular solutions (although it may be a challenge to construct inflating regions and de Sitter vacua~\cite{Hertzberg:2007ke}, see however~\cite{Silverstein:2007ac}). We hope to return to these and other issues in future work.

We thank T. Banks, S. Carroll, U.Danielsson, F. Denef, M. Dine, J. Preskill, A. Vilenkin, and K. Vyas for discussions and comments. MJ is supported by the Gordon and Betty Moore Foundation. ML was supported in part by the STINT CTP--Uppsala exchange, and thanks the CTP at MIT for hospitality.

\end{document}